\documentclass[aps,pre,amsmath,amssymb,preprint,longbibliography,%linenumbers,
runinaddress,
]{revtex4-1}
\ifx\pdfoutput\undefined
\usepackage{graphicx}
\else
\usepackage{graphicx}
\usepackage{epstopdf}
\fi
\usepackage[pdftitle={YX Huang},bookmarks=true,citecolor=red,colorlinks=false]{hyperref}
\usepackage{wasysym}
\usepackage{xcolor}
\usepackage{siunitx}

\usepackage{CJK}
\newcommand{\upd}{\mathrm{\,d}}

\newcommand{\red}[1]{\textcolor{black}{#1}}

\usepackage{chemformula,mhchem}

\begin{document}

%  \linenumbers
%\setpagewiselinenumbers
%\preprint{AIP/PoF}
\begin{CJK*}{GB}{gbsn} % Use default fonts from CJK (see below)
\title{Turbulent lithosphere deformation in the Tibetan Plateau}
% Force line breaks with %\\

\author{Xing Jian (¼òÐÇ)}
%\email{yongxianghuang@gmail.com}
\affiliation{State Key Laboratory of Marine Environmental Science, College of Ocean and Earth Sciences,
Xiamen University, Xiamen 361102,  China}

\author{Wei Zhang (ÕÅΡ)}
%\email{yongxianghuang@gmail.com}
\affiliation{State Key Laboratory of Marine Environmental Science, College of Ocean and Earth Sciences,
Xiamen University, Xiamen 361102,  China}

\
\author{Qiang Deng (µËÇ¿)}
%\email{yongxianghuang@gmail.com}
\affiliation{State Key Laboratory of Marine Environmental Science, College of Ocean and Earth Sciences,
Xiamen University, Xiamen 361102,  China}

\author{Yongxiang Huang (»ÆÓÀÏé)}
\email{yongxianghuang@gmail.com}
\affiliation{State Key Laboratory of Marine Environmental Science, College of Ocean and Earth Sciences,
Xiamen University, Xiamen 361102,  China}

%\author{Fengling Yu (Óà·ïÁá)}
%%\email{yongxianghuang@gmail.com}
%\affiliation{State Key Laboratory of Marine Environmental Science, College of Ocean and Earth Sciences,
%Xiamen University, Xiamen 361102,  China}

%\author{Nan Jiang (½ªéª)}
%\affiliation{Department of Mechanics, Tianjin University, 300072 Tianjin, China}
%
%
%\author{Yulu Liu (ÁõÓî½)}%
%%\email{francois.schmitt@univ-lille1.fr}
%\affiliation{Shanghai Institute of Applied Mathematics and Mechanics, Shanghai University,
%Shanghai 200072,  China}

\date{\today}% It is always \today, today,

\begin{abstract}
 In this work, we show that  the Tibetan Plateau deformation   demonstrates a turbulence-like statistics, e.g., spatial invariance cross continuous scales.  A dual-power-law behavior is evident to show the existence of two  possible conversation laws for the enstrophy-like  cascade on the range  $500\lesssim r\lesssim 2,000\,\si{km}$ and
  kinetic-energy-like cascade on the range $50\lesssim r\lesssim 500\,\si{km}$.  The measured second-order structure-function scaling exponents $\zeta(2)$ are similar with the counterpart of the Fourier scaling exponents observed in the atmosphere, where in the latter case the earth rotation is relevant. The turbulent statistics observed here for nearly zero Reynolds number flow  is  favor to be interpreted by the geostrophic turbulence theory. Moreover, the intermittency correction is recognized with an intensity to be close to the one of the hydrodynamic turbulence of  high Reynolds number turbulent flows, implying a universal scaling feature of very different turbulent flows.  Our results not only shed new light on the debate regarding the mechanism   of the Tibetan Plateau deformation, but also lead to new challenge for the geodynamic modelling using Newton or non-Newtonian model that the observed  turbulence-like features have to be taken into account.   \end{abstract}

%{Time series analysis}
%\pacs{02.50.Fz}{Stochastic analysis}
%\pacs{47.27.Gs}{Isotropic turbulence; homogeneous
%turbulence}
%\pacs{47.53.+n}{Fractals in fluid dynamics}
%\keywords{autocorrelation function, power law}%Use showkeys class option if keyword
\maketitle
\end{CJK*}

\section{Introduction}
The Tibetan plateau, usually referred to as the ``Roof of the World" expresses a double-thickened crust and stands at an average elevation of 5 \si{km} over a region of approximately 3 million \si{km^2}, see Fig.\,\ref{fig:location}\red{\,(a)}. Given the 
India-Eurasia collision and uplift of the plateau as the most significant geological events on the earth during Cenozoic time, Tibetan plateau has been 
widely regarded as an ideal field laboratory for understanding the geodynamic processes of continental collision, deformation and the interactions between 
uplift and global climate change \cite{Harrison1992Science,Raymo1992Nature,Molnar1993RG,An2001Nature}. However, how the Tibetan plateau 
deformed and grew remains highly controversial. Proposed hypotheses mainly include, 1) %underthrusting of Indian continental lithosphere beneath the plateau \cite{Powell1986EPSL}, 2) 
rigid plates or blocks northward propagating subduction and extrusion \cite{Tapponnier1982GEO,Tapponnier2001Science}, 2) convective removal of mantle lithosphere and rapid, continuous and entire deformation \cite{Molnar1993RG}, and 3) lower crustal flow rather than substantial upper crustal thickening contributes the plateau deformation and uplift \cite{Royden1997Science,Clark2000GEO}. These models are very creative and highly provocative, represent distinct driving mechanisms and kinematic descriptions of surface deformation, and thus have attracted considerable attention for decades. To test these hypotheses, a great number of geological and geophysical data and various methods have been used, primarily including paleoaltimetry, thermochronology, basin analysis and magnetostratigraphy, global positioning system (GPS) data and subsurface geophysical data analyses \cite{Zhang2004GEO,Rowley2006Nature,Clark2010EPSL,Bai2010NatGeo,Lease2012Tec,Hoke2014EPSL,Jian2018GR}. 
Although none of these models uniquely account for all of the geological and geophysical data and observations, more and more studies are aware of the presence of continuous medium and the important role of the rheology in the surface deformation of the Tibetan Plateau \cite{Clark2000GEO,Zhang2004GEO,Royden2008Science}. However, to the best of our knowledge, the  spatial scale invariance of such flowing deformation has never been taken into account.

Turbulence or turbulence-like phenomena are ubiquitous in the nature, which is often characterized by scale invariance in both spatial and temporal domains. It ranges from the evolution of the universe \citep{Gibson1996AMR}, movement of atmosphere and ocean \citep{Nastrom1984Nature,Thorpe2005book},  the painting  by  Leonardo da Vinci \citep{Frisch1995} or  van Gogh \citep{Aragon2008JMIV},  
  collective motion of bacteria \citep{Wensink2012PNAS,Qiu2016PRE},  the Bose-Einstein condensate \citep{Navon2016Nature}, \red{financial activity \citep{Ghashghaie1996a,Schmitt1999,Lux2001QF,Li2014PhysicaA,Mandelbrot2007}},  etc. 
Note that turbulence is usually recognized  by its main features that a broad range of spatial and temporal scales or many degrees of freedom are excited in the dynamical system  \cite{Groisman2000Nature,Biferale2018PR}. 
The turbulence theory is thus such theory to describe the energy
injection and dissipation patterns or the balance among  other physical quantities. 
This pattern could be quite different for different dynamical systems. For instance, in the classical
 three-dimensional hydrodynamical turbulence, the energy is injected into the system at large-scale and is transferred  to small-scale, and so on, until to the viscosity scale where the kinetic energy is converting to heat \cite{Frisch1995}. This is a forward energy cascade with the famous Kolmogorov $5/3$-law for the spatial Fourier 
 power spectrum of high Reynolds number turbulent flows, e.g., $E(k)\propto k^{-5/3}$. While in the two-dimensional 
 turbulence, the energy (resp. enstrophy, square of vorticity) is inputted into the system through a middle scale. It is 
 then transferred upward (resp. downward) due to energy (resp. enstrophy) conservation with a $5/3$-law for 
 large-scale part (resp. $3$ scaling law for small-scale part) \cite{Kraichnan1967PoF}.  Another famous example is the 
 theory of  geostrophic turbulence, in which the horizontal pressure gradient is balanced by the  Coriolis force \cite{Charney1971JAS}.  A potential enstrophy cascade with a scaling exponent $3$ (resp. large-scale part) and energy cascade with a scaling exponent $5/3$ (small-scale part)
are then presented \cite{Vallgren2011PRL}.

 In this work, in the spirit of the turbulence theory, we show that  the Tibetan Plateau deformation  also demonstrates a turbulence-like statistics, e.g., spatial invariance cross continuous scales.  A dual-power-law behavior is evident to show the existence of two possible conversation laws for the (potential) enstrophy-like  cascade on the range  
 $500\lesssim r\lesssim2,000\,\si{km}$ and kinetic energy-like cascade on the range $50\lesssim r\lesssim 500\,\si{km}$.  The measured second-order structure-function (SF) scaling exponents $\zeta(2)$ are similar with the ones observed in the atmosphere \cite{Nastrom1984Nature}, where in the latter case the earth rotation is relevant. The turbulent statistics observed here  is  favor  to be interpreted by the geostrophic turbulence theory, where a large-scale forcing  due to the India-Euraisa collision might be balanced by the Coriolis force. Furthermore, the intermittency correction is identified with a strength close to the one of three-dimensional hydrodynamical turbulence of high Reynolds number turbulent flows.
 Our results not only shed new light on the debate regarding the mechanism   of the Tibetan Plateau deformation, but  also lead to new challenge for the geodynamic modeling that the observed  turbulent features have to be taken into account.

\section{Data and Methodology}

\begin{figure*}[!htb]
\centering
 \includegraphics[width=0.85\linewidth]{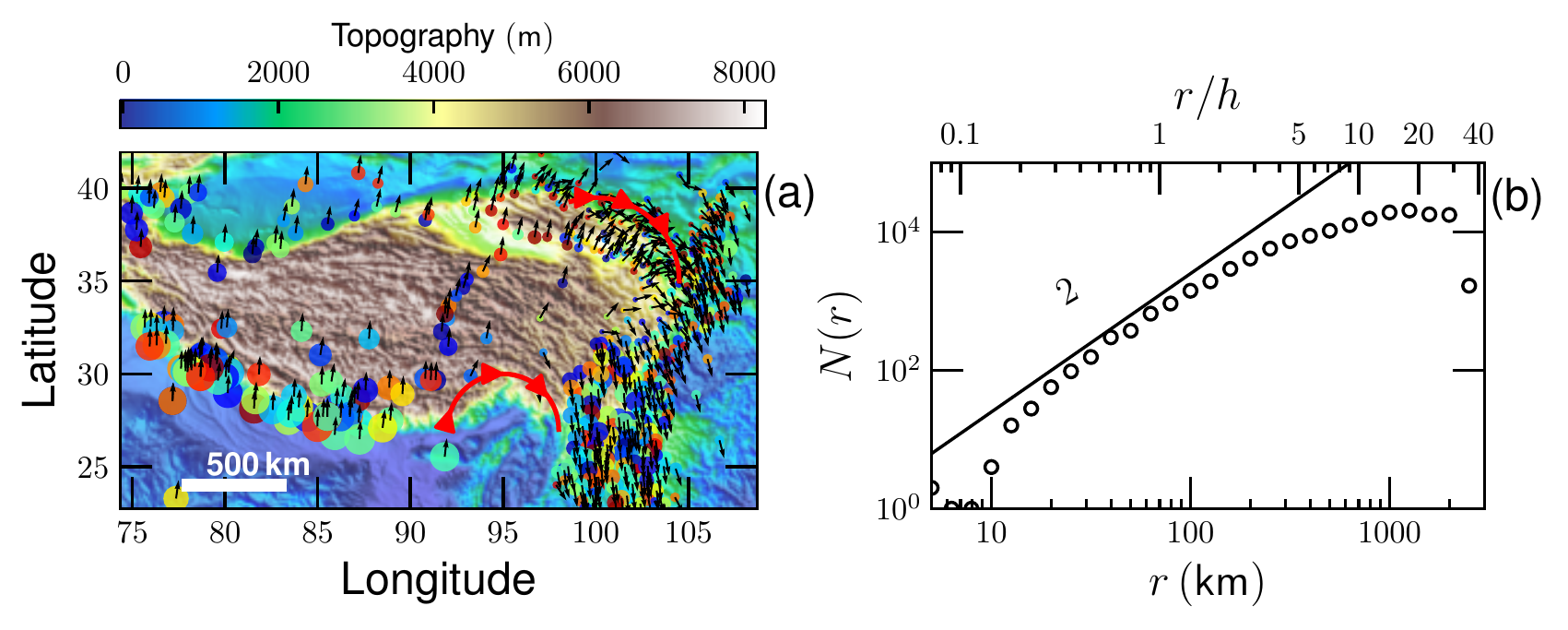}
  \caption{(Color online) (a) Spatial distribution of the 553 GPS monitoring stations at Tibetan Plateau, where the velocity amplitude in \si{cm} is encoded by symbol size. 
  The velocity unit vector \red{is} indicated by an arrow. Large clockwise rotation of crustal
material around the eastern Himalayan syntaxis are illustrated by \red{the big} arrow. The GPS velocity  data are taken from Ref.\,\cite{Zhang2004GEO}. The color map  is the elevation provided by ETOPO1 [35]. (b) The experimental number distribution of the neighbor distance,  where $r$ is the great circle distance and $h=70\,\si{km}$ is the average depth of the lithosphere. The solid line indicates a power-law relation with a scaling exponent $2$ for reference.   %%%%%%%
}\label{fig:location}
\end{figure*}

\begin{figure}[!htb]
\centering
 \includegraphics[width=0.85\linewidth]{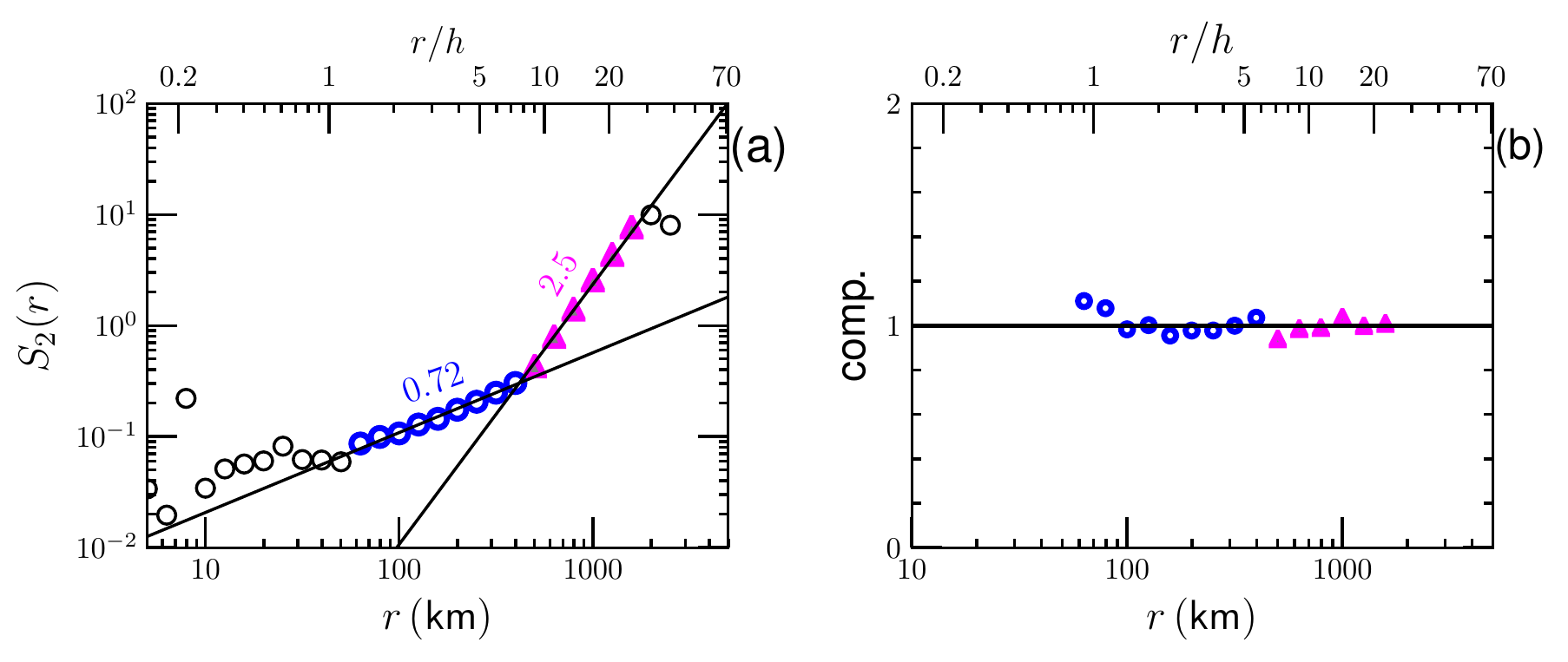}
  \caption{(Color online) (a)  Measured second-order structure-functions. The solid line is the fitting on the range $50\lesssim r\lesssim500\,\si{km}$ (resp. $0.7\lesssim r/h\lesssim7$) and $500\lesssim r\lesssim 2,000\,\si{km}$  (resp. $7\lesssim r/h\lesssim28$) with  respectively scaling exponent  $0.72\pm0.07$ and $2.50\pm0.07$. (b) The corresponding compensated curve, e.g., $S_2(r)r^{-\zeta(2)}C^{-1}$ using a fitted parameter, to highlight the power-law behavior.  
  }\label{fig:SF2}
\end{figure}

The GPS velocity data set is provided in Ref.\,\cite{Zhang2004GEO}. Figure \ref{fig:location} shows the  deformation velocity unit vector collected from 553 monitoring locations \cite{Zhang2004GEO}, where the topology provided by ETOPO1 \footnote{\href{https://www.ngdc.noaa.gov/mgg/global}{https://www.ngdc.noaa.gov/mgg/global}}  is illustrated in a color map. The symbols indicates the velocity magnitude in the range $0.17\sim 3.95\,\si{cm/year}$.  Their mean magnitude and  standard deviation respectively are  $1.27\,\si{cm/year}$ and $0.97\,\si{cm/year}$.
Figure \ref{fig:location}\,(b) shows the distribution of the neighbor distance of two pair of monitoring locations. Note that a power-law behavior with a scaling exponent $2$ indicates a homogeneous distribution of these monitoring stations, which is illustrated by a solid line for reference in Fig.\,\ref{fig:location}\,(b).  Roughly speaking, the monitoring stations are homogeneous distributed on the scale range $20\lesssim r\lesssim 200\,\si{km}$ (resp., $0.3\lesssim r/h\lesssim 3$, where $h=70\,\si{km}$ is the average depth of the Tibetan lithosphere).

The velocity pattern demonstrates an anticyclone (clockwise) structure, showing eddy-like motions. %This is consistent with the fact that the motion here might mainly due to the  India-Eurasia collision.
 To characterize the motions more quantitatively,  we introduce here a second-order moment of the structure-function (SF), which is written as,
 \begin{equation}
 S_2(r)=\langle  \vert \mathbf{u}(\mathbf{x}+\mathbf{r})-\mathbf{u}(\mathbf{x})\vert^2 \rangle
 \end{equation}
 where $r=\vert \mathbf{r}\vert$ is the great circle distance, $\mathbf{u}$ is the velocity vector.  For a scaling process, one expects the following relation, 
 \begin{equation}
 S_2(r)\propto r^{\zeta(2)}
 \end{equation}
 Figure \ref{fig:SF2}\,(a) shows the measured second-order SF $S_2(r)$. A dual-power-law behavior is evident respectively on the range $50\lesssim r\lesssim 500\,\si{km}$ (resp., $0.7\lesssim r/h\lesssim 7$) and $500\lesssim r\lesssim 2,000\,\si{km}$ (resp. $7\lesssim r/h\lesssim28$).
  The experimental scaling exponent are found to be $\zeta^\mathrm{S}(2)=0.72\pm0.07$  and $\zeta^\mathrm{L}(2)=2.50\pm0.07$, where the error indicates a $95\%$ fitting confidence level.  
According to the Wiener-Khinchin theorem, the Fourier power spectrum of the deformation velocity also follows a power-law behavior \citep{Frisch1995,Huang2010PRE}, 
\begin{equation}
E(k)\propto k^{-\beta}, \quad \beta=1+\zeta(2)
\end{equation}
Thus the scaling exponent of the Fourier power spectrum indicated by  second-order SF are $\beta^\mathrm{L}=3.5$ and $\beta^\mathrm{S}=1.72$, which could be verified in the future when more data are available from either observation or numerical simulation. 
 
 \section{Results and Discussions}
 
 \subsection{Scaling of deformation}

 The value of the scaling exponent $\zeta^{\mathrm{S}}(2)$ provided by the second-order SF
 is close to $2/3$, which implies a kinetic energy cascade that has been predicted by several theories. For example,  the Kolmogorov 1941 theory for three-dimensional homogeneous and isotropic turbulence  
   for the fully developed hydrodynamic turbulence \cite{Frisch1995};  the Kraichnan 1967 theory for 
   the two-dimensional turbulence \cite{Kraichnan1967PoF};  Charney 1971 theory for the 
   geostrophic  turbulence \citep{Charney1971JAS}. The another scaling value $\zeta^{\mathrm{L}}(2)$  may
 imply  a (potential) enstrophy  conservation  in the framework of two-dimensional turbulence \cite{Kraichnan1967PoF} or  geostrophic turbulence \cite{Charney1971JAS}.  
As aforementioned, the mechanism behind the power-law is  the pattern between injection and dissipation. Therefore, to exclude any possible explanations,  the external force that driving the lithosphere deformation has to be recognized.  A possible driving force is from the collision between the Indian 
and Eurasian plates with  a large-scale instability above $2000\,\si{km}$. Another possibility of the external force is at scale around $500\,\si{km}$, where the kinetic energy is injected into the system via the thermal plumes of the mental convection \cite{Zhong2005PoF}.
Due to the complexity of the current problem,  such balance pattern  is more complex than the ideal  2D turbulence theory or geostrophic turbulence theory. 
With the limited data, we cannot rule out any one of them.
A scale-to-scale energy/enstrophy flux  should be checked with attention to identify the cascade direction when the data is available \cite{Biferale2018PR,Zhou2015JFM}.

 It is interesting to note that a similar  dual-power-law behavior has been 
 reported for the atmospheric movement in the Fourier space with the same 
 separation scale around $500\,\si{km}$ around the same latitude \cite{Nastrom1984Nature,Gao2016AE}, where the separation scale $500\,\si{km}$ is determined by   geostrophic balance that described by  the Rossby number, see definition below.
 According to \citet{Vallgren2011PRL}, if the large-scale forcing due to the India-Eurasia collision  is applicable, then the geostropic turbulence is favorable. 
   Note that the power-law behavior observed here is consistent with the discovery of the continuous deformation  by \citet{Zhang2004GEO} for the spatial scale above $100\,\si{km}$.

%%%%%%%%%%%%%%%
 \begin{figure}[!htb]
\centering
\includegraphics[width=0.8\linewidth,clip]{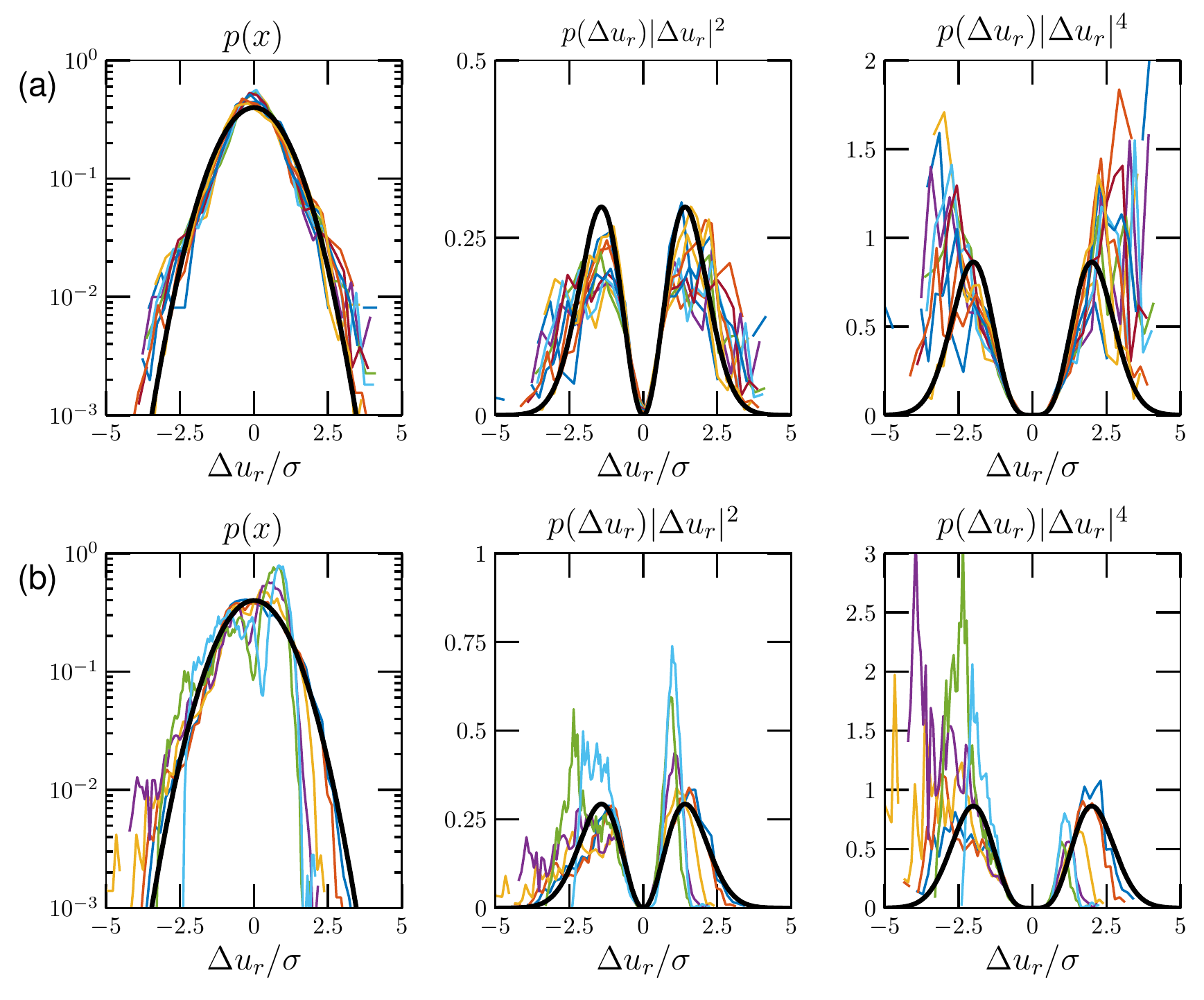}
  \caption{(Color online) Experimental probability density function $p(x)$ and the corresponding integral kernel $p(\Delta u_r)\vert \Delta u_r \vert^q$ for various separation scale $r$: (a) $50\lesssim r\lesssim 500\,\si{km}$, and (b) $500\lesssim r \lesssim 2,000\,\si{km}$\red{, where the thin line with different color indicates different separation scales, and the thick solid line is the normal distribution for reference.} Due to the finite sample size, the fourth-order structure-function is \red{slightly} deviating from statistical convergence.  
 }\label{fig:PDFs}
\end{figure}

The basic characteristic of turbulent system is intermittency, manifested as intense and sporadic fluctuations on different scale of motions.  It is  one of the most fascinating  feature of the hydrodynamic turbulence \cite{Frisch1995}, which has been reported also for other complex dynamic systems \cite{Schmitt2016Book}. 
To track  such intermittency correction, a high-order SFs is introduced, i.e.,
\begin{equation}
\red{ S_q(r)=\langle  \vert \mathbf{u}(\mathbf{x}+\mathbf{r})-\mathbf{u}(\mathbf{x})\vert^q \rangle \propto r^{\zeta(q)}}
\end{equation}
\red{where $\zeta(q)$ is the scaling exponents for high-order SFs. $\zeta(q)$ is linear with $q$ if there is no intermittency correction and vice versa. The deviation from linear relation is usually believed to be an effect of the nonlinear interactions between different scales \citep{Frisch1995}, manifesting as a large variation of the considered data \citep{Schmitt2016Book}.}
Note that the  $S_q(r)$ can be also defined via a $r-$dependent probability density function (pdf) $p(\Delta  u _r)$ of velocity difference $\Delta  u _r=\mathbf{u}(\mathbf{x}+\mathbf{r})-\mathbf{u}(\mathbf{x})$, 
\begin{equation}
S_q(r)=\langle \vert \Delta u_r \vert ^q \rangle=\int p(\Delta  u _r) \vert\Delta  u _r\vert^q\upd \Delta  u _r
\end{equation}
where $p(\Delta  u _r) \vert\Delta  u _r\vert^q$ is the $q$th-order integral kernel. \red{To check whether the statistics is convergent or not}, we plot the measured pdf and the corresponding integral kernel in Fig. \ref{fig:PDFs} for (a) $50\lesssim r \lesssim 500\,\si{km}$ and (b) $500\lesssim r\lesssim 2000\,\si{km}$. It suggests a safe estimation of the high-order SFs on the range $-1\le q\le 4$ with this limit data set.

\begin{figure}[!htb]
\centering
 \includegraphics[width=1\linewidth]{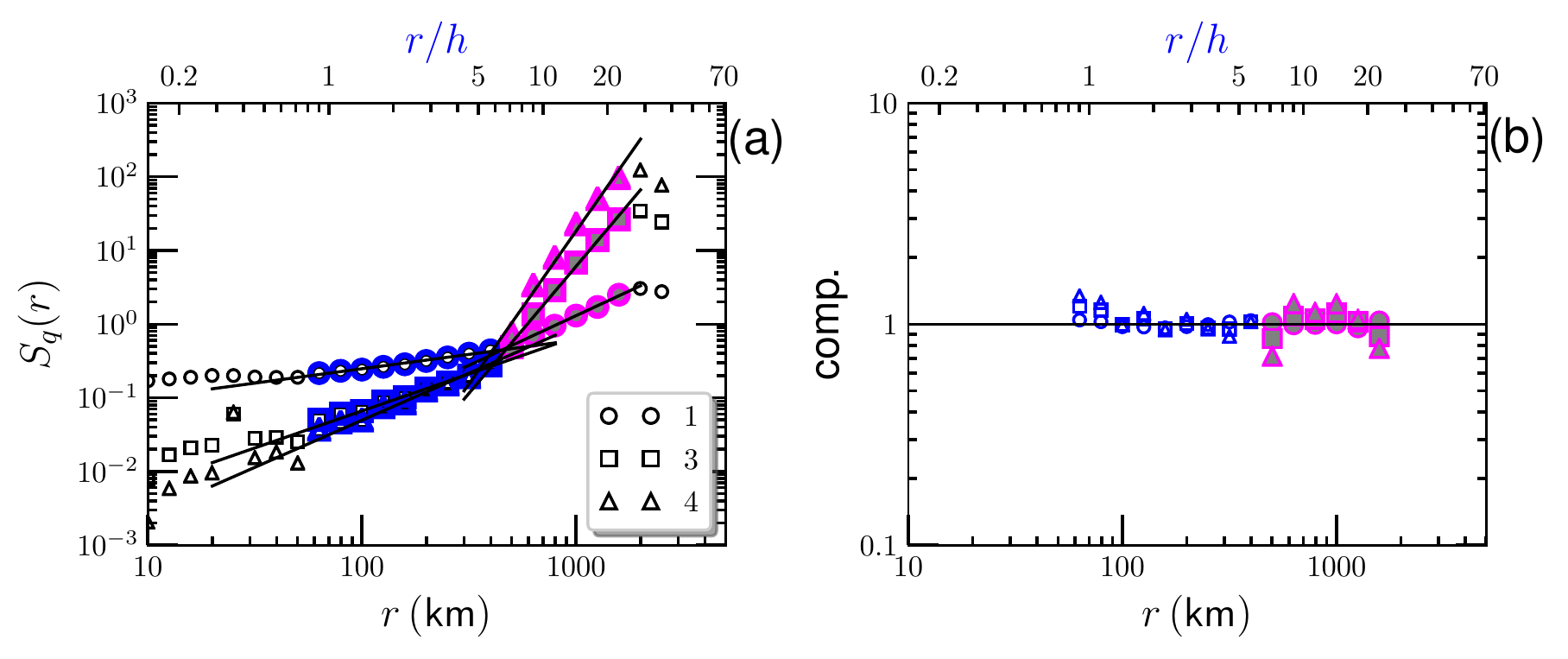}
  \caption{(Color online) (a) Measured high-order structure-functions $S_q(r)$ for $q=1,\,3,\,4$. The solid line illustrates the power-law fitting. (b) The corresponding compensated curve to highlight the power-law behavior. 
  }\label{fig:SFn}
\end{figure}

\begin{figure}[!htb]
\centering
 \includegraphics[width=1\linewidth,clip]{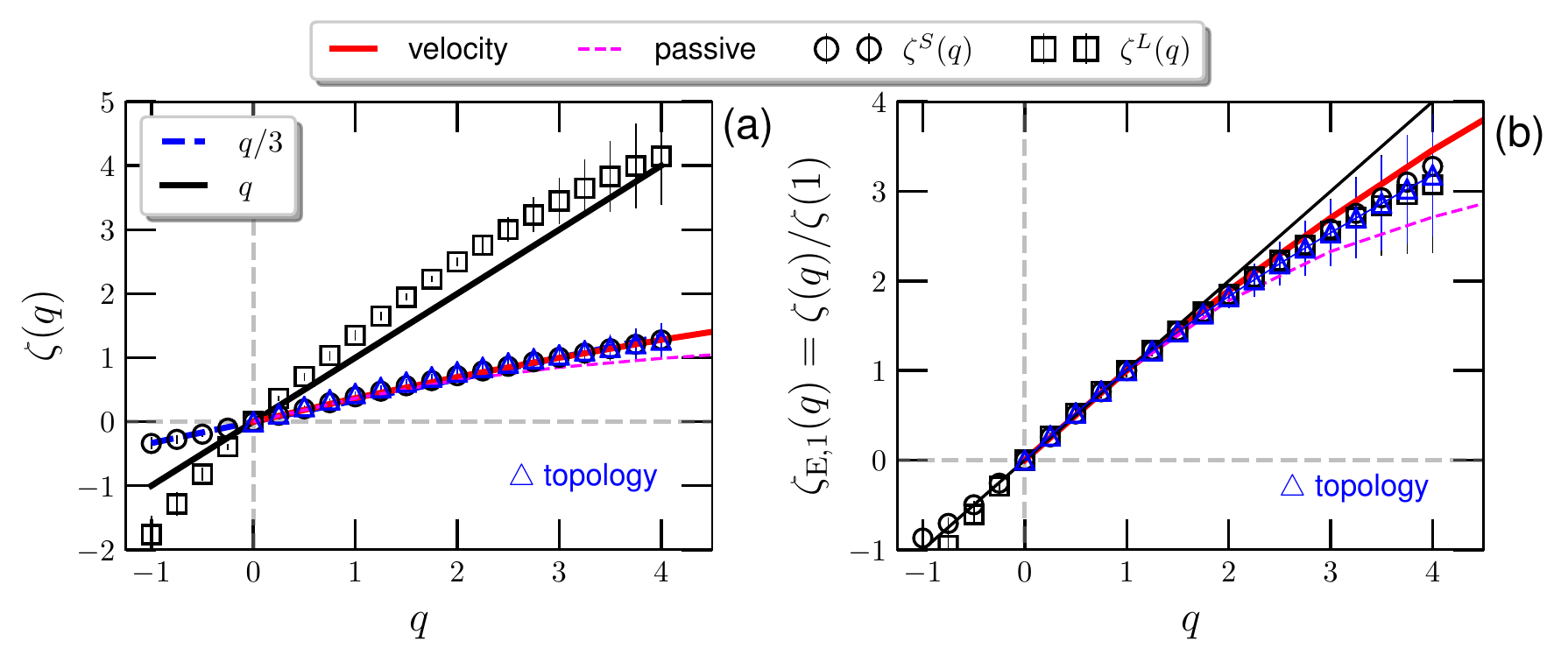}
  \caption{(Color online) (a) Measured high-order scaling exponent $\zeta(q)$ of small ($\ocircle$) and large ($\square$) scale motions for $-1\le q\le 4$, where $\zeta(q)=q/3$ (dashed line) and $\zeta(q)=q$ (solid line) are illustrated for reference.  (b) The extended-self-similarity plot of the measured scaling exponent $\zeta_\textrm{E,1}(q)=\zeta(q)/\zeta(1)$. For comparison, the scaling exponent compiled for velocity (thick solid line) \citep{Schmitt2006} and passive scalar (thin solid line) \citep{Schmitt2005}   are also shown. The errorbar indicates a 95\% fitting confidence level.  The scaling exponent for topology is illustrated as $\triangle$.}\label{fig:Scaling}
\end{figure}

High-order SFs
 are then calculated with $-1\le q\le 4$. However, only the case for $0\le q\le 4$ is discussed below. 
 Figure \ref{fig:SFn}\,(a) shows the SFs for $q=1$ ($\ocircle$), $3$ ($\square$) and $4$ ($\triangle$), where the solid line is a least square fitting. The dual-power-law is evident on the same scale ranges.  Figure \ref{fig:SFn}\,(b) shows the corresponding compensated curves to highlight  the observed power-law behavior.  Figure \ref{fig:Scaling}\,(a) shows the measured scaling exponents   
 for small-scale part $\zeta^\mathrm{S}(q)$ ($\ocircle$) and large-scale one $\zeta^\mathrm{L}(q)$ ($\square$). For comparison,  $\zeta(q)=q/3$ for the energy cascade  and $\zeta(q)=q$ for the (potential) enstrophy  cascade are also shown. First of all,  the experimental curves are convex, confirming the existence of intermittency correction.  
Secondly, the scaling  exponent $\zeta^{\mathrm{S}}(q)$ for the scale on the range $50\lesssim r \lesssim 500\,\si{km}$ is close to the value $\zeta(q)=q/3$, indicating an energy cascade with intermittency correction. Thirdly, the scaling exponent $\zeta^{\mathrm{L}}(q)$ for the scale on the range $500\lesssim r \lesssim 2,000\,\si{km}$ close to the value $\zeta(q)=q$, indicating a (potential) enstrophy cascade with intermittency correction. To characterize the intensity of intermittency, the extended-self-similarity \citep{Benzi1993PRE} is resorted via 
plotting $\zeta_{\mathrm{E,1}}(q)=\zeta(q)/\zeta(1)$ versus $q$, see  Fig.\,\ref{fig:Scaling}\,(b). The experimental curves 
$\zeta_{\mathrm{E,1}}^{\mathrm{S}}(q)$ and $\zeta_{\mathrm{E,1}}^{\mathrm{L}}(q)$ collapse with each other when $0\le q \le 3$. While the former one is slightly above the latter one when $3\le q \le 4$. For comparison, the scaling value of the hydrodynamical turbulence (thick solid line) \citep{Schmitt2006} and passive scalar turbulence (thin solid line) \citep{Schmitt2005} are also illustrated. Graphically, the measured scaling exponents are close to the one of hydrodynamical turbulence of high Reynolds number turbulent flows, implying a possible universal feature of very different turbulent systems \citep{Wang2017PRE}.

 %The external forcing could be at scale above 2000 km   due to the India-Eurasia collision or one at 500 km due to the mantle convection \cite{Zhong2005PoF}. 

\subsection{Scaling of topography}

\begin{figure}[!htb]
\centering
\includegraphics[width=0.85\linewidth,clip]{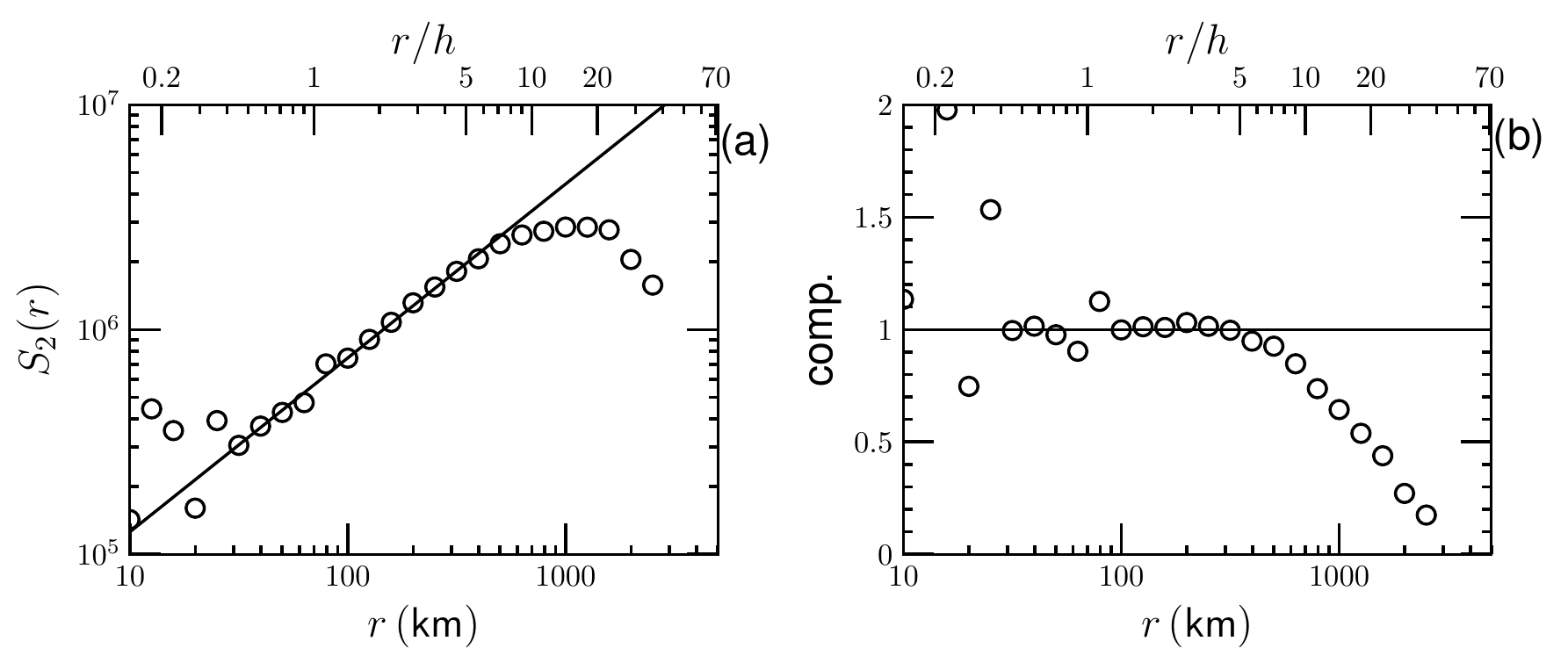}
  \caption{(Color online) a) Experimental second-order structure-function for the topography using the ETOPO1 data. Power-law behavior is observed on the range $50\lesssim r\lesssim 500\,\si{km}$ with a scaling exponent $0.77\pm0.07$. b) The corresponding  compensated curve using fitted parameters to highlight the power-law behavior.
 }\label{fig:TSFs}
\end{figure}

To cross verify the above observation, 
the topography of the Tibetan Plateau  provided by ETOPO1 % \footnote{\href{https://www.ngdc.noaa.gov/mgg/global}{https://www.ngdc.noaa.gov/mgg/global}}
 is analyzed below \citep{Gagnon2003EPL}.
 The evolution of the elevation can be approximately  written as, 
\begin{equation}
h(x,t)=\int_0^t v_h(x,t') dt'\simeq \tilde{v}_h(x)  t+h(x,0)
\end{equation}
where $h(x,t_0)$ is the initial elevation, and the typical vertical velocity  $\tilde{v}_h(x)$ can be treated as average vertical velocity  since it is a very slowly variation with time.
The elevation difference is thus an approximation of the velocity difference, 
%\begin{widetext}
\begin{equation}
\Delta_r h(x,t) \propto\Delta_r  \tilde{v}_h (x)  t
%\simeq \frac{\left\{v_h(x+r)-v_h(x)\right\}\Delta t+h(x+r,t_0)-h(x,t_0)}{\Delta t}%=\Delta v_r
\end{equation}
%\end{widetext}
where  $h(r+x,0)-h(x,0)\ll \left\{\tilde{v}_h(x+r)-\tilde{v}_h(x)\right\} t$ is assumed when $r$ is smaller than a certain value, e.g. $500\,\si{km}$. 
High-order SFs, e.g., $S_q(r)=\langle \vert \Delta _r h\vert^q  \rangle$ are estimated. A single power-law behavior is observed on the scale range $50\lesssim r \lesssim 500\,\si{km}$ (resp. $0.7\lesssim r/h\lesssim 7$) in Fig.\,\ref{fig:TSFs}\,(a). The measured second-order SF scaling exponent  $\zeta(2)$ is to be $\zeta(2)=0.77\pm0.07$, which agrees well with the one of the horizontal deform velocity obtained on the same scale range. Figure \ref{fig:TSFs}\,(b) shows the corresponding compensated curve to emphasize the observed scaling behavior. 
The measured high-order scaling exponent $\zeta(q)$ with $0\le q\le 4$ is also shown in Fig.\,\ref{fig:Scaling} as $\triangle$. Graphically, it agrees well with the one of the horizontal velocity, confirming the existence of the turbulence-like dynamics.

\subsection{Several key parameters}

Finally, several possible relevant parameters are discussed as following. 
A scale dependent Reynolds number can be defined as, 
\begin{equation}
\mathrm{Re}(r)=\frac{\rho ur}{\mu}
\end{equation}
where $\rho$ is the density,  $u$  the  velocity, $r$ the spatial scale and $\mu$ is the dynamic viscosity. It characterizes the ratio between the inertia and viscosity forces.
Typical values in Tibet are $\rho\simeq 3 \,\si{g/cm^3}$  \citep{Li2017GG}, $u\simeq 1.27 \,\si{cm/year}$ \citep{Zhang2004GEO}, $\mu=10^{19}\sim 10^{24}\,\si{Pa\cdot s}$ \citep{Shi2008ESF}, respectively.  Considering the spatial scale $r$ from $50\,\si{km}$ to $2,000\,\si{km}$, one has a Reynold number on the range $2.4\times 10^{-24}\lesssim \mathrm{Re}(r)\lesssim 6\times 10^{-21}$, suggesting that the viscosity force is more relevant  than the inertial one in the current problem. \red{Note that turbulence is usual associated with the high-Reynolds number flows, where the inertia of fluid is relevant and the viscosity force can be neglected \citep{Frisch1995}. 
Turbulence without inertia \citep{Larson2000Nature} have been reported for the bacterial turbulence   \citep{Wensink2012PNAS,Qiu2016PRE,Wang2017PRE}, elastic turbulence \citep{Groisman2000Nature,Groisman2001Nature}, etc., with nearly zero Reynolds number. 
As aforementioned, the balance pattern  could be quite different from the classical hydrodynamic turbulence. However, they might share the universal turbulent features, such as the same strength of the intermittency \citep{Wang2017PRE}.}

Due to the earth rotation, the Coriolis force could be important. A scale-dependent Rossby number is introduced to characterize this effect, which is written as,
\begin{equation}
\mathrm{Ro}(r)=\frac{u}{2\Omega \sin(\phi)r}
\end{equation}
where  $2\Omega\sin(\phi)$ is the so-called Coriolis frequency, and $\Omega=7.27\times 10^{-5}\,\si{rad/s}$  the angular frequency of planetary rotation and $\phi\simeq 30\,\si{degree}$ the latitude.  It measures the ratio  between the inertial force and Coriolis force. 
Taking the same typical velocity and length scales as \red{for} the Reynolds number, one has typical  value of $\mathrm{Ro}(r)$ on the range $2.77\times10^{-12} \lesssim \mathrm{Ro}(r)\lesssim 1.11\times10^{-10}$, implying that the Coriolis force is more relevant than the inertial one.

%Ekman number is defined as,
%\begin{equation}
%\mathrm{Ek}(r)=\frac{\nu}{2r^2\Omega\sin(\phi)}
%\end{equation}
%where $\nu=\mu/\rho$ is the kinematic viscosity.  One  has $1.83\times 10^{10}\lesssim \mathrm{Ek}(r)\lesssim1.15\times10^{12}$.

Finally, a scale-dependent Deborah number is defined as,
\begin{equation}
\mathrm{De}(r)=\frac{t_{\mathrm{c}}(r)}{t_{\mathrm{p}}}
\end{equation}
where $t_{\mathrm{c}}(r)$ refers to the stress relaxation time for a given spatial scale $r$, and $t_{\mathrm{p}}$ is the time scale of observation.
It characterizes the ratio of the relaxation time $t_{\mathrm{c}}(r)$ characterizing the time it takes for a material to adjust to applied stresses or deformations, and the characteristic time scale $t_{\mathrm{p}}$ of an experiment  probing the response of the material.
 At lower Deborah numbers, the material behaves in a more fluid-like manner, with an associated Newtonian viscous flow. At higher Deborah numbers, the material behavior enters the non-Newtonian regime, increasingly dominated by elasticity and demonstrating solid-like behavior. Typical examples are flows of ice-river, asphaltum, etc., that for a long time observation and thus a small Deborah number, they behave in the  fluid-like manner \cite{Reiner1964}.  For the current case, the Deborah number is estimated on the range $0\le \mathrm{De}(r)\ll1 $, suggesting that the lithosphere deformation can be treated as fluid flow.

\section{Conclusion}

In summary,  in this work, the lithosphere deformation of the Tibetan  Plateau is analyzed in the spirit of the multiscale statistics from turbulence community. The dual-power-law behavior is evident respectively on the range $50\lesssim r \lesssim 500\,\si{km}$ (resp. $0.7\lesssim r/h\lesssim 7$) and $500\lesssim r \lesssim 2,000\,\si{km}$ (resp. $7\lesssim r/h\lesssim 28$). The scaling feature of the former scaling range indicates an energy cascade, while the one of  the latter scaling range implies a (potential) enstrophy cascade, which might be interpreted in the framework of geostrophic turbulence since one possible external force can be identified from the India-Eurasia collision with a spatial scale above $2,000\,\si{km}$.  The similar multiscale feature with the one from atmosphere suggests that they might share the similar  dynamic, e.g.,  the balance between large-scale force and Coriolis force. However, to exclude any  other possibility, a scale-to-scale energy/enstrophy flux has to be estimated either in Fourier space or physical domain to determine the cascade direction \citep{Biferale2018PR,Zhou2015JFM}, see example in Ref.\,\citep{Wang2017PRE}.  Moreover, the intermittency is revealed via the high-order SFs. With the help of extended-self-similarity, the intensity of intermittency is found to be the same as the one of hydrodynamic turbulence of high Reynolds turbulent flows, showing a universal feature of very different turbulent flows, even for nearly zero Reynolds number flows \citep{Groisman2000Nature,Groisman2001Nature,Qiu2016PRE,Wang2017PRE}. 
Our results not only shed new light on the understanding of the lithosphere deformation of Tibetan Plateau, but also lead to new challenge for the geophysical modelling using Newtonian or non-Newtonian fluid model that the observed turbulent features have to be taken into account.

 \begin{acknowledgments}
This work is sponsored by the National Natural Science Foundation of China (under Grant No. 11732010 and  41806052), and  partially by the Fundamental Research Funds for the Central Universities (Grant No.  20720180120, 20720180123 and 20720170073) and MEL (State Key Laboratory of Marine Environmental Science) Internal Research Fund (Grant No. MELRI1802). 
  We thank professor J.~Q. Zhong and F.~G. Schmitt for useful discussion.  The source code and GPS data   is available at: \footnote
 {\href{https://github.com/lanlankai}{https://github.com/lanlankai}}. \red{We thank the two anonymous reviewers for their careful reading and insightful comments and suggestions.}

\end{acknowledgments}

%\newpage
 %\bibliographystyle{apsrev}
%\bibliographystyle{aipnum4-1}
%\bibliography{all}% Produces the bibliography via BibTeX.

\end{document}